\documentclass{PoS}
\usepackage{cite}

\title{Baryon production from small to large collision systems at ALICE}

\ShortTitle{Baryon production from small to large collision systems at ALICE}

\author{\speaker{Omar V\'{a}zquez Rueda}{ on behalf of the ALICE Collaboration}\\
        Lund University\\
        E-mail: \email{omar.vazquez@hep.lu.se}}

\abstract{Studies of the production of light- and heavy-flavor baryons are of prominent importance to characterize the partonic phase created in ultrarelativistic heavy-ion collisions and to investigate hadronization mechanisms at the LHC. Studies performed in p--Pb and pp collisions have revealed unexpected features, qualitatively similar to what is observed in larger collision systems and, in the charm sector, not in line with the expectations from ${\rm e}^{+}{\rm e}^{-}$ and ${\rm e}^{-}{\rm p}$ interactions. The ALICE experiment has exploited its excellent tracking and particle identification capabilities down to low transverse momentum to perform an extensive study of protons, hyperons and charmed baryons. In this paper, a discussion of the most recent results on light (protons and hyperons) and heavy-flavor $(\Lambda_{\rm c})$ baryon production is presented, together with a comparison to phenomenological models.\\}

\FullConference{
European Physical Society Conference on High Energy Physics - EPS-HEP2019 -\\
			10-17 July, 2019\\
			Ghent, Belgium}

\begin{document}

\section{Introduction}
\noindent 
Measurements of ultrarelativistic heavy-ion collisions at the top LHC energies have corroborated the formation of a strongly coupled partonic state of matter, the so-called Quark--Gluon Plasma (QGP). Some of the features of this system are strong collective radial and anisotropic flow and opacity to jets~\cite{QGP}. Collective radial flow in central collisions is observed as hardening of the transverse momentum $(p_{\rm T})$ spectra of heavy particles such as protons while parton energy loss comes out as the suppression of the production of high-$p_{\rm T}$ particles~\cite{PbPb 5TeV}. Recent measurements in high-multiplicity pp and p--Pb collisions have revealed flow-like patterns even in these small collision systems~\cite{pp vs mult 7TeV,p-Pb 5TeV}. Furthermore, it has been observed that the yields of strange hadrons normalized to the one of pions $(\pi^{+}+\pi^{-})$ increase significantly with the charged-particle multiplicity of the event~\cite{strangeness paper}. This increase scales with the strangeness content of baryons. In high-multiplicity events, strangeness production reaches values similar to those observed in heavy-ion collisions, where a QGP is formed.\\
In this paper, a brief review of selected results of baryon production is reported. The results include both the multiplicity and $p_{\rm T}$-dependent baryon-to-meson ratios in pp, p--Pb and Pb--Pb collisions in the light and heavy-flavor sectors, as well as comparisons with model predictions. Measurements of the yields of strange hadrons relative to the one of pions across energy and system size are also discussed. The paper is organized as follows. In section 2, the ALICE apparatus and some of the techniques used to measure the $p_{\rm T}$ spectra of identified particles are described. Results and discussions are presented in section 3. Final remarks are summarized in section 4. 

\section{ALICE apparatus}
\noindent
ALICE (A Large Ion Collider Experiment) is the dedicated heavy-ion experiment at the LHC with unique capabilities for tracking and particle identification (PID) over a wide range of $p_{\rm T}$. The trigger selection and event classification into multiplicity classes are accomplished using the V0 detector. It is composed of a pair of forward scintillator hodoscopes, which cover the pseudorapidity ranges $2.8<\eta<5.1$ (V0A) and $-3.7<\eta<-1.7$ (V0C)~\cite{ALICE apparatus}. The V0 detector signals, which, are proportional to the charged-particle multiplicities, are used to define multiplicity classes for all three collision systems.\\

\noindent
The main detectors for tracking and PID of the central barrel are located inside a solenoidal magnet providing a magnetic field of 0.5 T. The innermost detector is the Inner Tracking System (ITS)~\cite{ALICE apparatus}, which is composed of six layers of silicon detectors covering the pseudorapidity range $|\eta| \leq 0.9$ and the full azimuthal range. In addition, it is also used for triggering and vertex reconstruction. The  PID techniques with the ITS rely on the specific energy loss $({\rm d}E/{\rm d}x)$ with a resolution of about 6\%. The Time Projection Chamber (TPC)~\cite{ALICE apparatus} consists of a hollow cylinder with a symmetry axis parallel to the beam axis. The active volume is about $90~{\rm m}^{3}$ filled with a mixture of gas consisting of ${\rm Ne}$-${\rm CO}_{2}$-${\rm N}_{2}$ at atmospheric pressure. An electrode at the center of the cylinder provides, together with a voltage dividing network at the surface of the outer and inner cylinder, a drift field of $400~{\rm V/cm}$. Its acceptance covers the pseudorapidity interval $|\eta| < 0.9$ and full azimuthal angle. Charged particle identification in the TPC relies on their ${\rm d}E/{\rm d}x$ in the gas volume\iffalse An analysis of the production of pions, kaons and (anti-)protons based on the ${\rm d}E/{\rm d}x$ in the relativistic rise region of the TPC can provide a measurement of the $p_{\rm T}$ spectra from $\sim 2~{\rm GeV}/c$ up to $20~{\rm GeV}/c$ \fi. The  Time-Of-Flight (TOF)~\cite{ALICE apparatus} detector is a large area array of Multigap  Resistive Plate Chambers (MRPC), positioned at $370-399~{\rm cm}$ from the beam axis and covering the full azimuthal angle and the pseudorapidity range $|\eta|<0.9$. The TOF detector can provide information about the production of identified particles in the intermediate $p_{\rm T}$ region, from $\sim 1~{\rm GeV}/c$ up to $3~{\rm GeV}/c$ for pions, kaons and (anti-)protons. Strange hadrons such as ${\rm K}_{\rm S}^{0}$ and $\Lambda(\overline{\Lambda})$ are reconstructed at mid-rapidity $(|y|<0.5)$ via their characteristic weak decay topologies: ${\rm K}_{\rm S}^{0} \rightarrow \pi^{+}+\pi^{-}$, $\Lambda(\overline{\Lambda}) \rightarrow {\rm p}(\bar{{\rm p}}) + \pi^{-}(\pi^{+})$. The $\Lambda^{+}_{\rm c}$ and $\overline{\Lambda}^{-}_{\rm c}$ are measured by reconstructing the hadronic decay modes: $\Lambda^{+}_{\rm c} \rightarrow {\rm p}{\rm K}^{-}\pi^{+}$ and $\Lambda^{+}_{\rm c} \rightarrow {\rm p}{\rm K}^{0}_{\rm S}$ (and charge conjugates).\\

\section{Results and discussions}
\begin{figure}[!h]
\centering
        \includegraphics[width=0.8\textwidth]{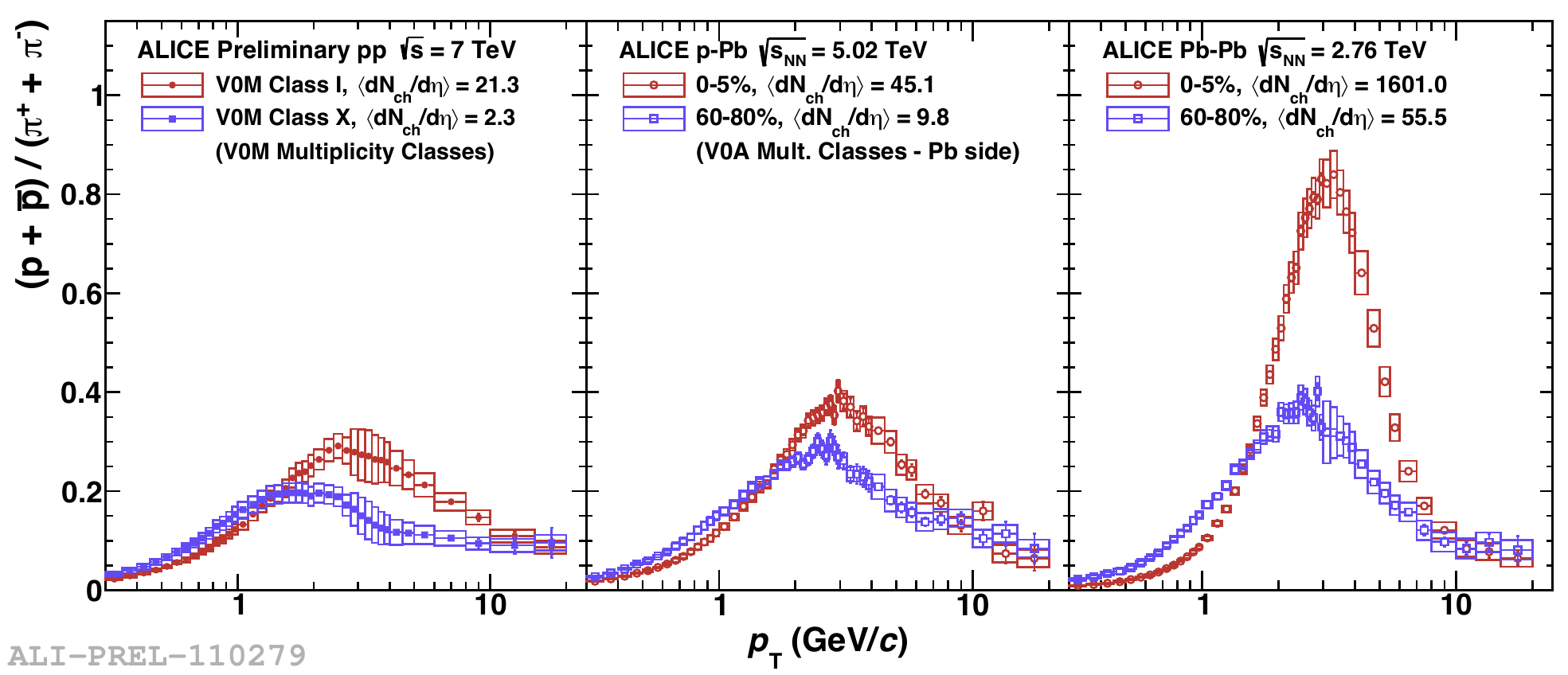}\\
        \includegraphics[width=0.8\textwidth]{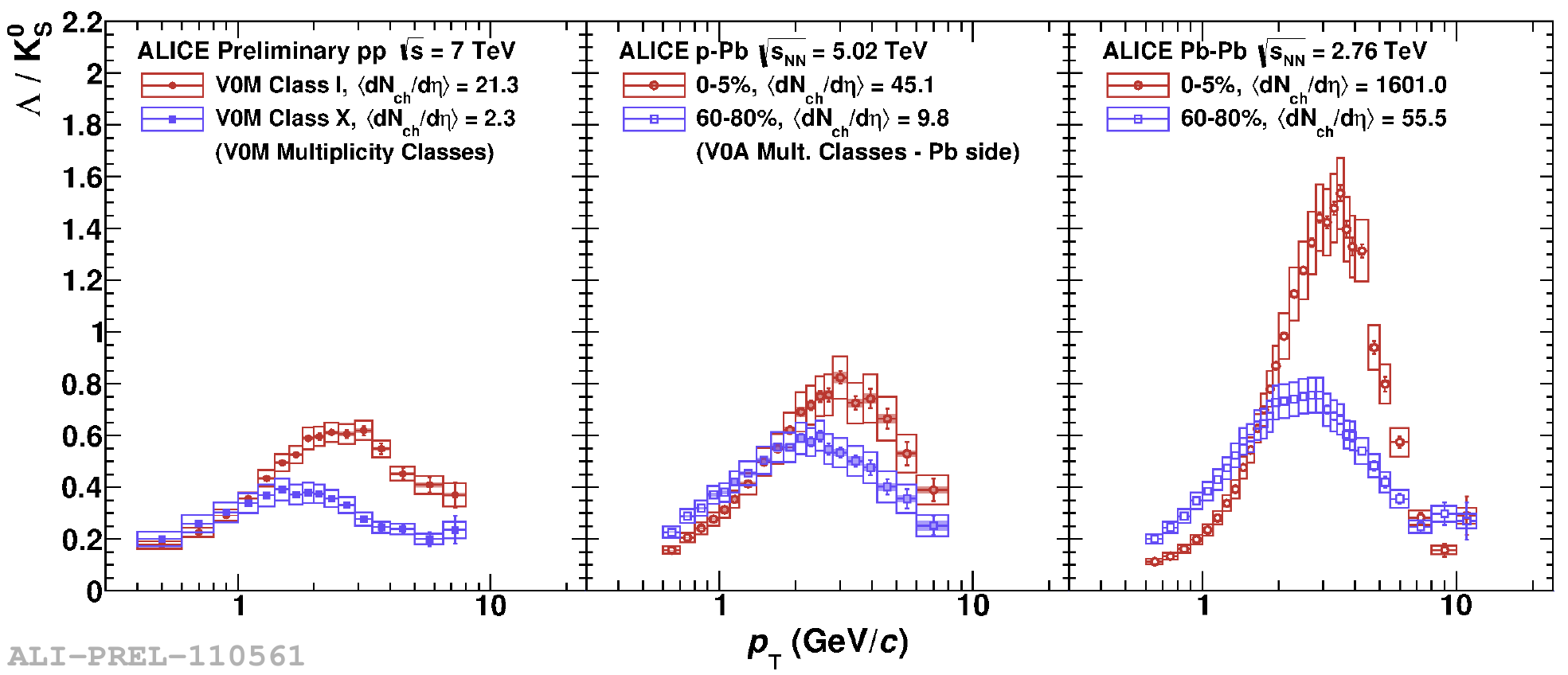}
\caption{$p_{\rm T}$-differential proton-to-pion (top) and $\Lambda/{\rm K}^{0}_{\rm S}$ (bottom) ratios. From the left to the right subpanel in each of the figures, the results from pp, p--Pb and Pb--Pb collisions are shown. Two event multiplicity classes are shown: red (blue) markers represent the results from the highest (lowest) multiplicity events in the respective colliding systems. The error bars show the statistical uncertainty, while the empty boxes shows the total systematic uncertainty.}
\label{fig:baryon-to-meson ratio}
\end{figure}

\noindent
The study of particle production as a function of multiplicity can be performed by calculating the ratios of particle $p_{\rm T}$ spectra to that of pions. Pions are commonly taken as the reference species since they are the most abundant particles and least affected by collective flow. Figure~\ref{fig:baryon-to-meson ratio} shows the $p_{\rm T}$-differential proton-to-pion and $\Lambda/{\rm K}^{0}_{\rm S}$ ratio in pp, p--Pb and Pb--Pb collisions. The results between the lowest and highest event multiplicity classes in each of the systems are contrasted. For the three different colliding systems the same pattern is observed: At low-$p_{\rm T}$ the high-multiplicity ratio is lower than the low-multplicity ratio, but this changes at high $p_{\rm T}$. This suggests that particles heavier than pions are shifted from low to high $p_{\rm T}$ as the charged particle multiplicity increases. While the maximum in the proton-to-pion ratio reaches an approximate value of $0.4$ in high multiplicity p--Pb collisions, the one from the lambda-to-pion ratio in the same multiplicity class reaches a value close to $0.8$. This is interpreted in terms of collective radial flow, whose effects are more relevant for heavier particles. Moreover, the largest value of the maximum seen in central Pb--Pb collisions reveals the presence of the strongest radial flow as discussed in terms of a Blast-Wave analysis in~\cite{PbPb 5TeV}.\\

\begin{figure}[!h]
\centering
        \includegraphics[width=0.6\textwidth]{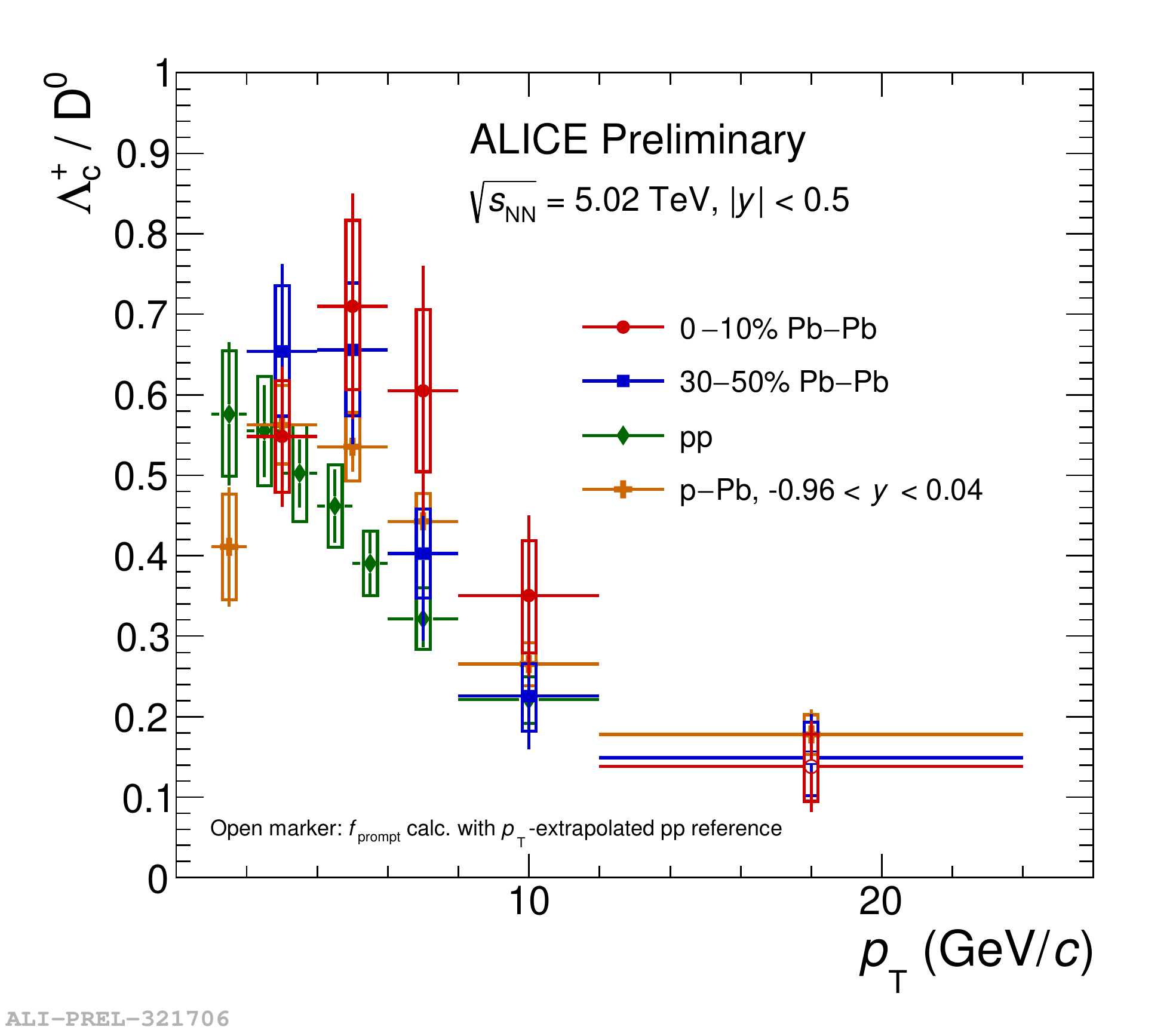}
\caption{$p_{\rm T}$-differential $\Lambda_{\rm c}^{+}/{\rm D}^{0}$ ratio in pp, p--Pb, central and peripheral Pb--Pb collisions at $\sqrt{s_{\rm NN}} = 5.02~{\rm TeV}$ measured at mid-rapidity $(|y|<0.5)$. The error bars show the statistical uncertainty, while the empty boxes show the total systematic uncertainty.}
\label{fig:Lambda-C-ratio PbPb}
\end{figure}

\noindent
The ALICE Collaboration has also measured the $\Lambda_{\rm c}^{+}/{\rm D}^{0}$ baryon-to-meson ratio in pp, p--Pb~\cite{Lambda charm} and Pb--Pb collisions at different energies. Due to the large mass of the charm quark $(m_{\rm c} \cong 1.3~{\rm GeV}/c^{2})$, it is produced in hard-partonic scattering processes during the initial stages of the collision. In Pb--Pb collisions they are even produced before the formation of the QGP, allowing them to interact with and experience the whole evolution of the medium. Figure~\ref{fig:Lambda-C-ratio PbPb} shows the $p_{\rm T}$-differential $\Lambda_{\rm c}^{+}/{\rm D}^{0}$ ratio in pp, p--Pb, central and peripheral Pb--Pb collisions at $\sqrt{s_{\rm NN}} = 5.02~{\rm TeV}$. It is observed that in central Pb--Pb collisions the ratio is higher than in peripheral ones at intermediate $p_{\rm T}$ $(4 \lesssim  p_{\rm T} \lesssim 10~{\rm GeV}/c)$. The result from central collisions shows a maximum at $p_{\rm T}\approx 5~{\rm GeV}/c$, which is consistent with the one measured in the proton-to-pion ratio of central Pb--Pb collisions although at a smaller value of $p_{\rm T}$. At low-$p_{\rm T}$ $(p_{\rm T} \lesssim 4~{\rm GeV}/c)$ a hint of suppression is observed for central collisions, however, this has to be treated with care as the systematic uncertainties are too large to draw firm conclusions. Also, the analogous measurements from pp and p--Pb collisions at intermediate $p_{\rm T}$ are found to be smaller than in Pb--Pb collisions. The enhancement of the baryon-to-meson ratios in the light-flavor and heavy-flavor sector at intermediate $p_{\rm T}$ suggest that the baryon production in an environment with a high density of partons occurs by recombination as discussed in~\cite{recombination 1}.\\ 

\begin{figure}[!h]
\centering
        \includegraphics[width=0.6\textwidth]{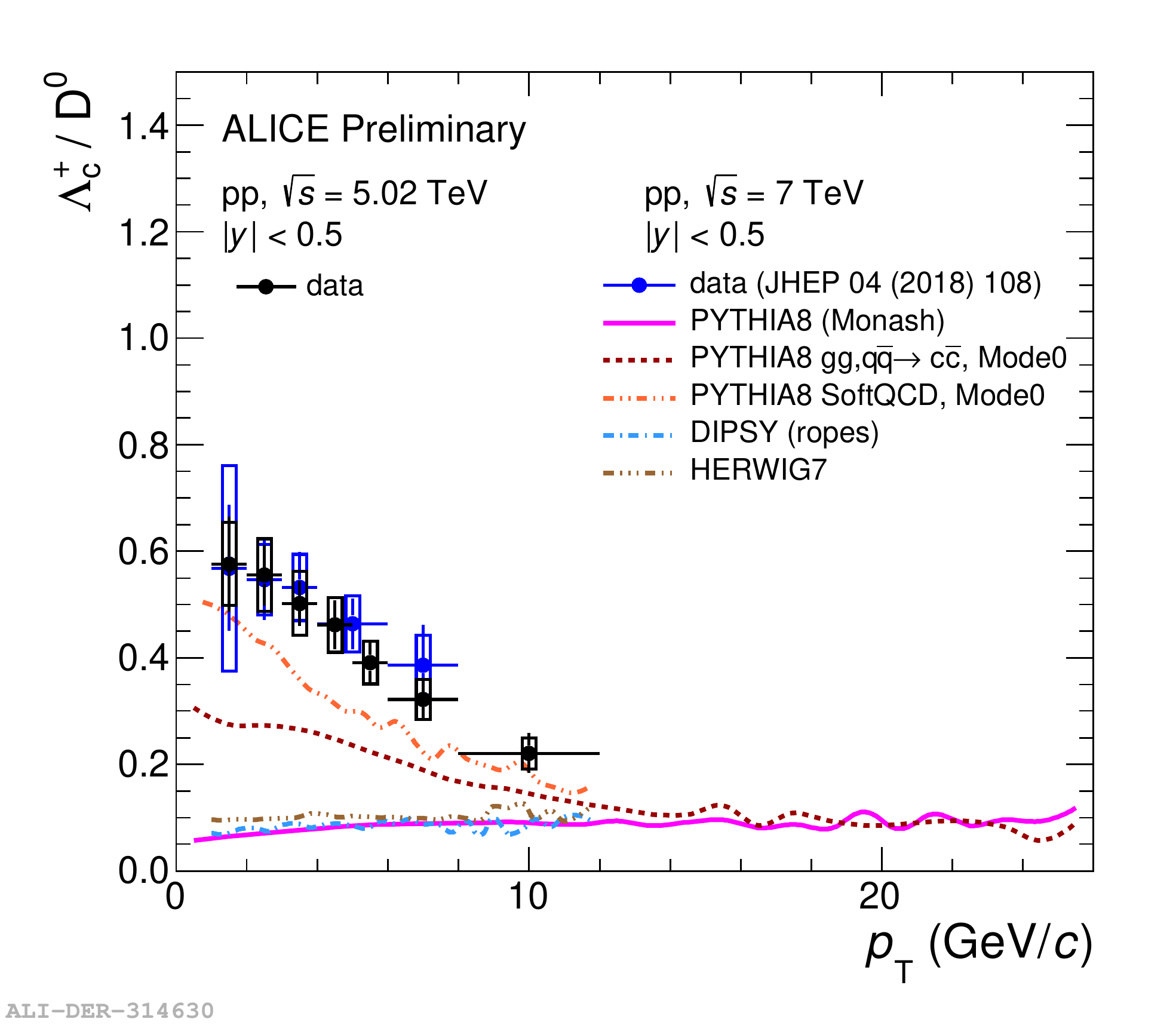}
\caption{$p_{\rm T}$-differential $\Lambda_{\rm c}^{+}/{\rm D}^{0}$ ratio in pp collisions at $\sqrt{s}=5.02~{\rm TeV}$ and at $\sqrt{s}=7~{\rm TeV}$ compared to predictions from MC generators. The error bars show the statistical uncertainty, while the empty boxes show the total systematic uncertainty.}
\label{fig:Lambda-C-ratio-light-heavy flavors}
\end{figure}

\noindent
Figure~\ref{fig:Lambda-C-ratio-light-heavy flavors} shows the $p_{\rm T}$-differential $\Lambda_{\rm c}^{+}/{\rm D}^{0}$ ratio in pp collisions at $\sqrt{s}=5.02~{\rm TeV}$ and at $\sqrt{s}=7~{\rm TeV}$ as well as predictions from MC generators and the expectations from ${\rm e}^{+}{\rm e}^{-}$ collisions. The measured values at the two different energies are consistent within uncertainties. The PYTHIA8 predictions include the Monash tune and modes where color reconnection can lead to baryon junctions. The baryon junction formation in PYTHIA brings the predictions closer to data. What remains to be seen is if one can also find a tune in which the junction mechanism also can describe correctly the light-flavor baryon production - in particular for protons, which are not enhanced in high-multiplicity events.   The model parameters corresponding to Mode0 used in these comparisons are described in~\cite{PYTHIA paper}. All the MC generators including the expectations from ${\rm e}^{+}{\rm e}^{-}$ were found to significantly underestimate the data at low $p_{\rm T}$ while at high $p_{\rm T}$ data seem to tend to the values predicted by these models. The color reconnection mechanism in PYTHIA enhances flow-like effects, bringing the predictions closer to data.\\ 

\begin{figure}[!h]
\centering
        \includegraphics[width=0.5\textwidth]{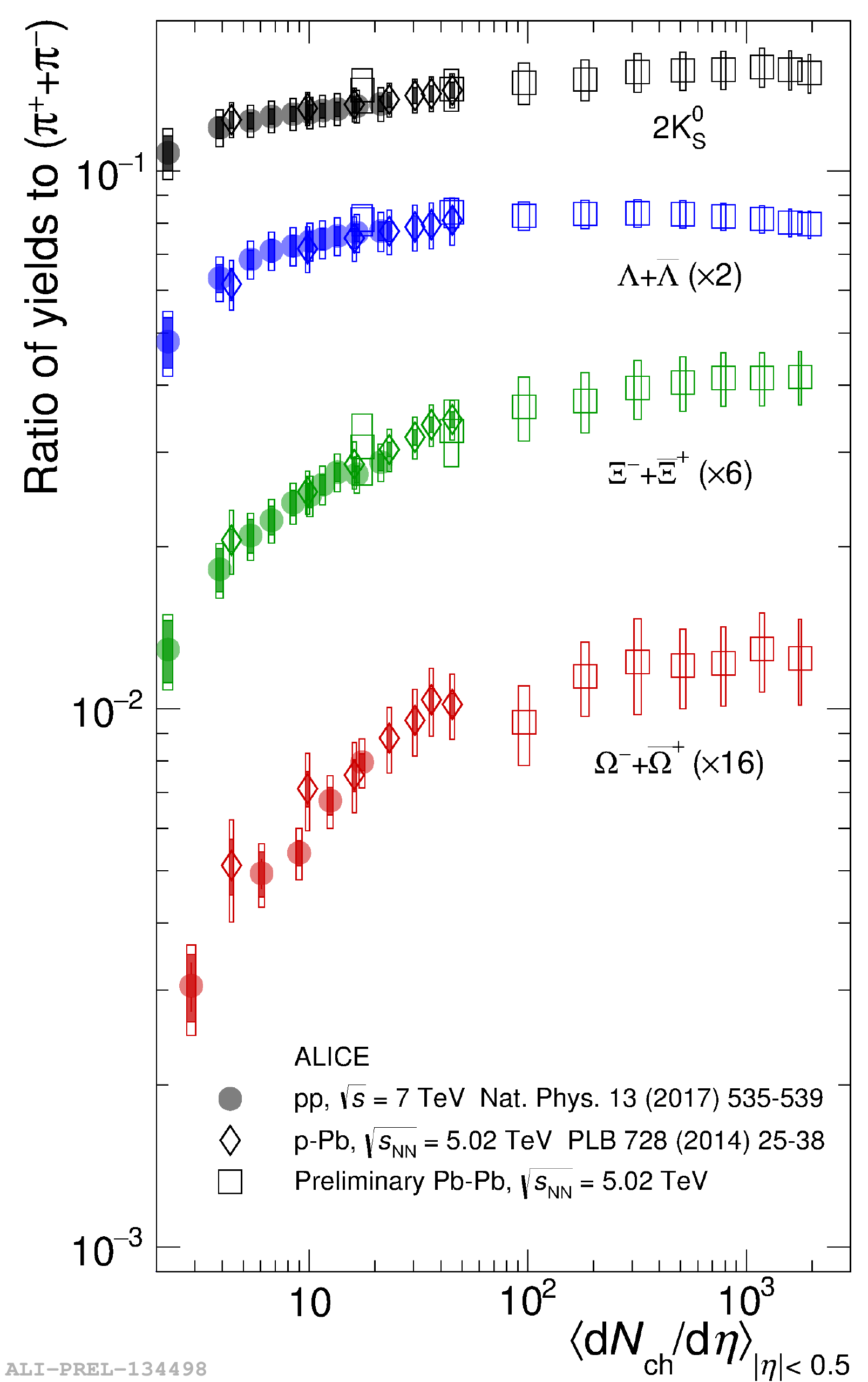}
\caption{$p_{\rm T}$-integrated yield ratios to pions $(\pi^{+}+\pi^{-})$ as a function of $\langle {\rm d}N_{\rm ch}/{\rm d}\eta \rangle$. Results from pp at $\sqrt{s}=7~{\rm TeV}$, p--Pb at $\sqrt{s_{\rm NN}}=5.02~{\rm TeV}$ and Pb--Pb collisions at $\sqrt{s_{\rm NN}}=2.76~{\rm TeV}$ are shown. The error bars show the statistical uncertainty, whereas the empty and dark-shaded boxes show the total systematic uncertainty and the contribution uncorrelated across multiplicity bins, respectively. }
\label{fig:strangeness ratios}
\end{figure}

\noindent
The yields of ${\rm K}^{0}_{\rm S}, \Lambda, \Xi~{\rm and}~\Omega$ normalized to the one of pions are shown in Fig.~\ref{fig:strangeness ratios} as a function of the mean charged-particle density $(\langle {\rm d}N_{\rm ch}/{\rm d}\eta \rangle)$ in pp at $\sqrt{s}=7~{\rm TeV}$, p--Pb at $\sqrt{s_{\rm NN}}=5.02~{\rm TeV}$ and Pb--Pb collisions at $\sqrt{s_{\rm NN}}=2.76~{\rm TeV}$. Significant enhancement of strange hadrons with respect to pions $(\pi^{+}+\pi^{-})$ with increasing multiplicity is observed. Moreover, the strangeness production increases with increasing strange-quark content. The origin of strangeness production in hadronic collisions is apparently driven by the characteristics of the final state rather than by the collision system or energy~\cite{strangeness paper}. Predictions from MC generators were compared and it was found that while PYTHIA8~\cite{PYHTIA8} and EPOS LHC~\cite{EPOS} underestimate the yield ratios, DIPSY~\cite{DIPSY} describes the data best. DIPSY is a model where interactions among strings allow the formation of `color ropes' which are expected to produce more strange hadrons than strings, since the `ropes' have stronger color fields (higher string tension).\\   

\section{Conclusions}
\noindent
The ALICE experiment has made precise measurements of spectra of identified particles down to low $p_{\rm T}$ in pp, p--Pb and Pb--Pb collisions over a broad range of collision energies allowing the exploration of the non-perturbative QCD regime. It has been observed that the light-flavor baryon-to-meson ratios in high multiplicity pp and p--Pb collisions showed qualitative similarities to the ones in heavy-ion collisions and that similar effects are also present in the heavy-flavor sector. Furthermore, precise measurements of the production of identified-charged particles as a function of multiplicity have shown that multiplicity is a key variable for studying the relative particle abundances.\\

\end{document}